\documentclass[prl,aps,twocolumn,floats,nofootinbib,showpacs]{revtex4}
\usepackage{amsmath,amssymb,graphicx,psfrag}

\begin{document}
\renewcommand{\ni}{{\noindent}}
\newcommand{\dprime}{{\prime\prime}}
\newcommand{\be}{\begin{equation}}
\newcommand{\ee}{\end{equation}}
\newcommand{\bea}{\begin{eqnarray}}
\newcommand{\eea}{\end{eqnarray}}
\newcommand{\la}{\langle}
\newcommand{\ra}{\rangle}
\newcommand{\dg}{\dagger}
\newcommand\lbs{\left[}
\newcommand\rbs{\right]}
\newcommand\lbr{\left(}
\newcommand\rbr{\right)}
\newcommand\f{\frac}
\newcommand\e{\epsilon}
\newcommand\ua{\uparrow}
\newcommand\da{\downarrow}
\title{Doping a correlated band insulator: A new route to half metallic behaviour}
\author{Arti Garg$^{1}$, H. R. Krishnamurthy$^{2}$ and Mohit Randeria $^{3}$}
\affiliation{$^{1}$Theoretical Condensed Matter Physics Division, Saha Institute of Nuclear Physics, 1/AF Bidhannagar, Kolkata 700 064, India \\
$^{2}$ Centre for Condensed Matter Theory, Department of Physics, Indian Institute of Science, Bangalore 560 012,
	and JNCASR, Jakkur, Bangalore 560 064, India\\
$^{3}$Department of Physics, The Ohio State University, Columbus, OH 43210,USA}
\vspace{0.2cm}
\begin{abstract}
\vspace{0.3cm}
We demonstrate in a simple model the surprising result that turning on
an on-site Coulomb interaction $U$ in a doped band insulator leads to
the formation of a half-metallic state. In the undoped system, we show
that increasing U leads to a first order transition between a
paramagnetic, band insulator and an antiferomagnetic Mott insulator at
a finite value $U_{AF}$. Upon doping, the system exhibits half
metallic ferrimagnetism over a wide range of doping and interaction
strengths on either side of $U_{AF}$. Our results, based on dynamical
mean field theory, suggest a novel route to half-metallic behavior and
provide motivation for experiments on new materials for spintronics.
\vspace{0.1cm}
\end{abstract}
\maketitle
Turning on strong electron correlations in a normal metallic system is generally believed to result in interesting phases like anti-ferromagnetic Mott Insulator, high $T_c$ superconductor, pseudogap phase and non fermi-liquid phases. But the effect of e-e interactions in a band insulator have not been explored in detail so far.
In this paper we study effects of onsite interaction $U$ on a band insulator and present a {\it {novel interaction driven route to half-metal phase}}. We show that doping a correlated band insulator results in the formation of half-metallic (HM) ferrimagnet.
HMs are an interesting class of materials in which electrons with one spin direction behave as in a metal and electrons with the opposite spin direction behave as in an insulator, and have applications in spintronics as they can generate spin-polarized currents~\cite{HMAFM_expt}.

 Specifically, in this paper we study a simple tight-binding model with two bands, arising from a staggered potential $\Delta$ on the sites of a bipartite lattice,  in the presence of an on-site Coulomb repulsion, the Hubbard $U$.
At half filling, when one band is filled and the other is empty, this system is a paramagnetic band insulator (BI). When $U$ is turned on, as we show, an anti-ferromagnetic (AFM) order sets in with a first order phase transition at some threshold $U=U_{AF}$. Upon doping, this system becomes a {\it {ferrimagnetic HM}} over a range of doping and $U$ values.

\begin{figure}
\begin{center}
\includegraphics[width=3.5in,angle=0]{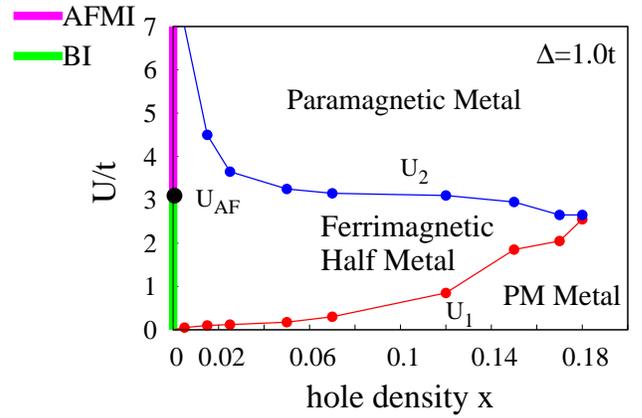}
\caption{
Zero temperature phase diagram of the model in Eq.~(\ref{model}) obtained within DMFT for the Bethe lattice of infinite connectivity. Circles show the data points and the lines are guides to the eye. At half filling, the system is a band insulator at weak $U$ and becomes an AFMI with a first order phase transition at $U_{AF}$.
Upon doping, it becomes a ferrimagnetic half-metal (HM) over a large range of doping and $U$ values.}

%Inset shows $U_{AF}$ vs $\Delta$ at half filling. At half filling a HM AFM point is seen at $U=U_{HM}> U_{AF}$.}

\label{phase_diag}
\end{center}
\vskip-6mm
\end{figure}

Intuitively the formation of a HM upon doping can be understood as follows.
Due to the staggered potential, the band gaps for the two spin components in the anti-ferromagnetic insulator (AFMI) phase are different, e.g. $Eg_{\da} < Eg_{\ua}$. On hole doping, in a rigid band picture, one would expect that it will be energetically favourable to put all the holes in the down spin band. This will make the down spin band conducting while the up-spin will remain insulating, resulting in a HM phase.  Similarly for $U < U_{AF}$, consider the BI in presence of a small staggered magnetic field $h \rightarrow 0$ such that the band gap in the single particle excitation spectrum of the two spin components is different, being $Eg_{\ua} = Eg+h$ and $Eg_{\da} = Eg-h$. Following the argument mentioned above, doping this BI with holes of density $x$ will result in the formation of a HM with a net moment $\sim x$ which, due to a molecular field arising from U, will self consistently cause the staggered magnetisation to be non zero.
The simple rigid band picture used in this argument will hold only for small doping and weak coupling regime. The question of interest is whether the HM phase exists on finite amount of doping at intermediate and large $U$ values. In this paper we show that it is possible to get the HM phase for a finite doping over a range of $U$ values as shown in the phase diagram of Fig.~\ref{phase_diag}.
In the HM phase, there occurs a full redistribution of the degrees of freedom as compared to the half filled case as shown in detail in later sections. The HM phase survives for the widest range of doping in the intermediate coupling regime where $U \sim 2\Delta$.

\begin{figure}[h!]
\begin{center}
\includegraphics[width=3.0in,angle=0]{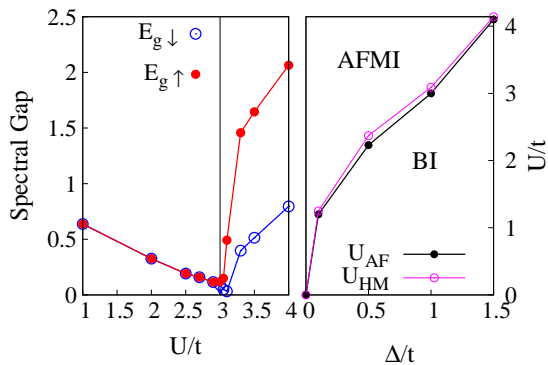}
\caption{Left panel: Spin-resolved spectral gaps $E_{g\ua}$ and $E_{g\da}$ plotted as a function of $U/t$ for $\Delta =1.0t$ at half-filling. For $U<U_{AF}$, in the BI phase, $E_{g\ua}=E_{g\da}$ and the gaps decrease with increasing $U/t$. At $U=U_{AF}$, there occurs a jump separating the two gaps such that $E_{g\da} < E_{g\ua}$.
For $U > U_{HM}$, in the AFMI phase, both the gaps increase with increasing $U/t$.
Right panel: $T=0$ phase diagram at half-filling. For $U < U_{AF}$, the system is a PM BI. At $U_{AF}$, a first order transition occurs with the onset of an AFM order. A HM AFM point is seen at $U=U_{HM}> U_{AF}$. For all $U > U_{HM}$, the system is an AFMI.}
\label{Gap}
\end{center}
\end{figure}

The model we consider has tight-binding electrons moving on a bipartite
lattice (sub-lattices A and B) described by
\bea
H=-t\sum_{i\in A,j\in B,\sigma} [~c^{\dagger}_{i\sigma}c_{j\sigma}
+h.c~]+ \Delta \sum_{i\in A}n_{i} -\Delta\sum_{i \in B} n_{i} \nonumber \\
%\mbox{~~~~~~~~~~~~~~~~~~~~~~~~~~~}
+U\sum_{i}n_{i\ua}n_{i\da}-\mu\sum_{i}n_{i}
\label {model}
\eea
where $t$ is the nearest neighbor
hopping, $U$ the Hubbard repulsion and $\Delta$ a staggered one-body
potential which doubles the unit cell. The chemical potential $\mu$ is fixed so that the
average occupancy is $\left(\langle n_A \rangle + \langle n_B \rangle \right)/2=n=1-x$. The Hamiltonian (\ref
{model}) is sometimes called the ``ionic Hubbard model'' (IHM) with
$\Delta$ the ``ionic'' potential.
In an earlier  work, using a dynamical mean field theory (DMFT) approach employing iterated perturbation theory (IPT) as the impurity solver, we studied this model at half filling in the spin-symmetric case and showed how strong correlations dynamically close the gap in a band insulator resulting in an intermediate metallic phase~\cite{garg}. This result was reproduced qualitatively by many other groups~\cite{ihm-metal}. Here we study this model using the same DMFT approach in the spin asymmetric case ~\cite{note}.
Once we allow for the magnetic order,  the system at half-filling undergoes a first order phase transition from a PM BI to AFMI (see Fig.~\ref{Gap}).
In the spin asymmetric phase there are some earlier works at half filling~\cite{afm_ihm} and for the doped case~\cite{doped} but to the best of our knowledge the existence of a HM phase has not been suggested so far.

The DMFT approximation is exact in the limit of large
dimensionality \cite{georges,jarrell} and has been demonstrated to be quite
successful in understanding the metal-insulator transition
\cite{georges,jarrell} in the usual Hubbard model, which is the
$\Delta =0$ limit of eq.~(\ref{model}). We focus in this paper on
the anti-ferromagnetic sector of
eq.~(\ref{model}), for which it is convenient to introduce the
matrix Green's function \be
\hat{G}_{\alpha\beta}^{\sigma}({\bf{k}},i\omega_n)
= \lbr \begin{array} {cc} \zeta_{A\sigma}({\bf{k}},i\omega_n) & -\epsilon_{\bf{k}} \\
-\epsilon_{\bf{k}} & \zeta_{B\sigma}({\bf{k}},i\omega_n)
\end{array}\rbr^{-1} \label{Greensfn} \ee where $\alpha,\beta$ are
sub-lattice ($A,B$) indices, $\sigma$ is the spin index, ${\bf k}$ belongs to the first Brillouin Zone (BZ) of \emph{one sub-lattice},
$i\omega_n=(2n+1)\pi T$ and $T$ is the temperature.
The kinetic energy is described by the dispersion
$\epsilon_k$ and $\zeta_{A(B)\sigma} \equiv i\omega_n \mp
\Delta+\mu-\Sigma_{A(B)\sigma}(i\omega_n)$. Within the DMFT approach the
self energy is purely local~\cite{georges}. Thus the diagonal
self-energies $\Sigma_{\alpha\sigma}(i\omega_n)$ are ${\bf
k}$-independent and the off-diagonal self-energies vanish
(since the latter would couple the A and B sub-lattices).
The DMFT approach includes {\it local} quantum fluctuations by
mapping \cite{georges,jarrell} the lattice problem onto a
single-site or ``impurity'' with local interaction $U$
hybridizing with a self-consistently determined bath as follows.
(i) We start with a guess for $\Sigma_{\alpha\sigma}(i\omega_n)$, $n_{\alpha\sigma}$, and compute the local
$G_{\alpha\sigma}(i\omega_{n})=\sum_{{\bf{k}}}G_{\alpha\alpha\sigma}({\bf{k}},i\omega_{n})$
rewritten as \be
G_{\alpha\sigma}(i\omega_n)=\zeta_{\bar{\alpha}\sigma}(i\omega_n)\int_{-\infty}^{\infty}
d\epsilon
\frac{\rho_{0}(\epsilon)}{\zeta_{A\sigma}(i\omega_n)\zeta_{B\sigma}(i\omega_n)-\epsilon^{2}}
\label{fullG} \ee with $\alpha=A(B)$,$\sigma=\ua,\da$ and $\bar{\alpha}=B(A)$ with $\rho_{0}(\epsilon)$ is the bare DOS for the lattice considered (see below).
(ii) We next determine the ``host Green's function"~\cite{georges,jarrell} $\mathcal{G}_{0\alpha\sigma}$ from the Dyson equation
$\mathcal{G}_{0\alpha\sigma}^{-1}(i\omega_n) = G_{\alpha\sigma}^{-1}(i\omega_n) + \Sigma_{\alpha\sigma}(i\omega_n)$. (iii)
We solve the impurity problem to obtain
$\Sigma_{\alpha\sigma}(i\omega_{n}) = \Sigma_{\alpha\sigma}
\left[\mathcal{G}_{0\alpha\sigma}(i\omega_{n})\right]$
(iv) We iterate steps (i), (ii) and (iii) till a self-consistent solution is obtained.
We use as our ``impurity solver'' in step (iii) a generalization
of the iterated perturbation theory (IPT) \cite{georges,kk} scheme
which has the merit of giving semi-analytical results directly in
the real frequency domain. %Details of generalised IPT are discussed in the supplementary material.

 For simplicity, here we present the results for the $T=0$ solution of
DMFT equations on a Bethe lattice of connectivity $z \rightarrow
\infty$. The hopping amplitude is rescaled as $t\rightarrow t/\sqrt{z}$ to
get a non-trivial limit and the bare DOS is then given by
$\rho_0(\epsilon) = \sqrt{4t^2-\epsilon^2}/(2\pi t^2)$ which
greatly simplifies the integral in eq.~(\ref{fullG}). The phase diagram in Fig.~\ref{phase_diag} has been obtained from an analysis of various physical quantities which we describe in detail below. We believe that the results obtained will be qualitatively similar in case of other generic compact DOS. In the discussion below, first we describe the detailed results for the half-filled case followed by the results for the doped case.
\begin{figure}[h!]
\begin{center}
\vskip-1.0cm
\hspace{-1.8cm}
\includegraphics[width=1.7in,angle=-90]{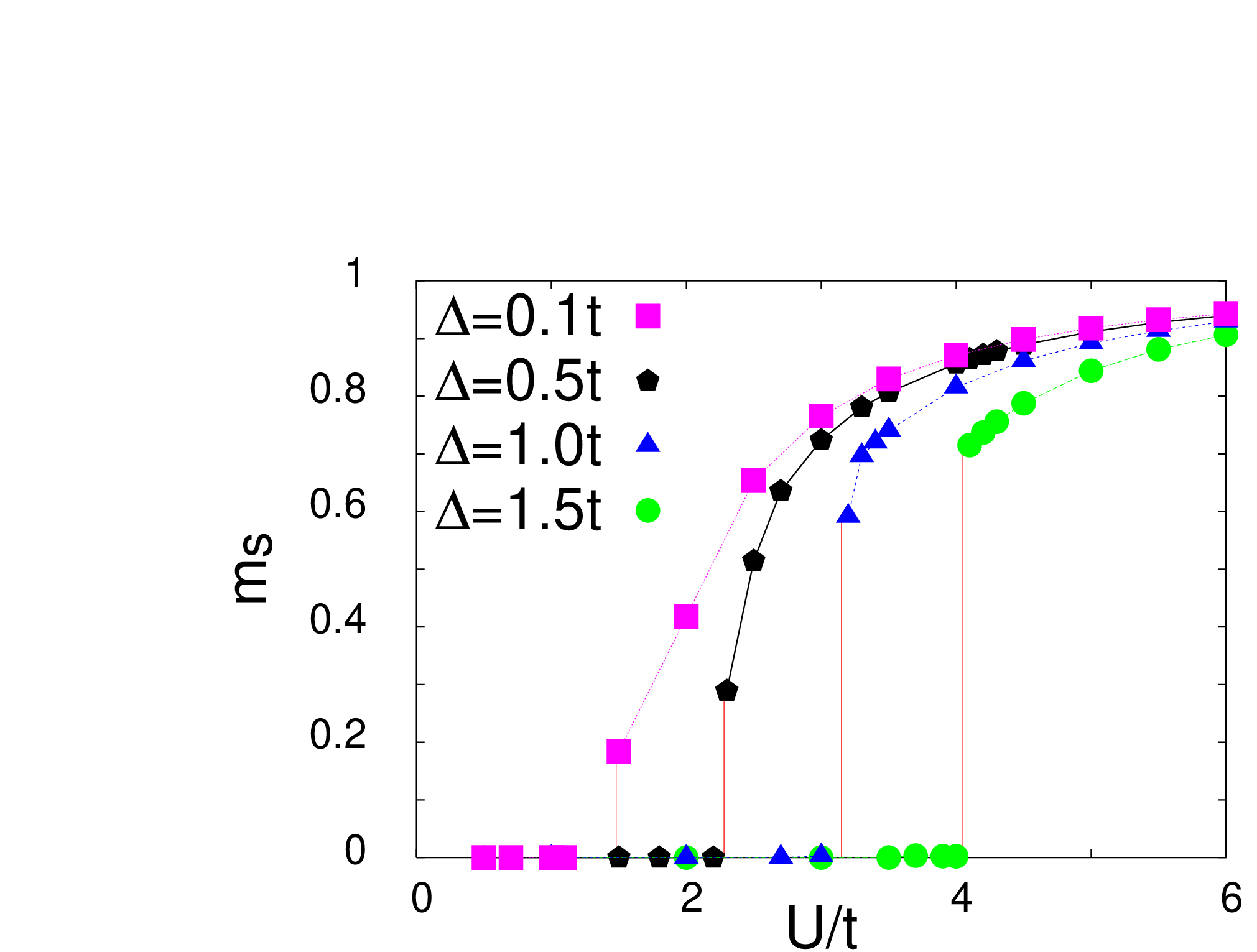}
\hspace{-2.25cm}
\includegraphics[width=1.7in,angle=-90]{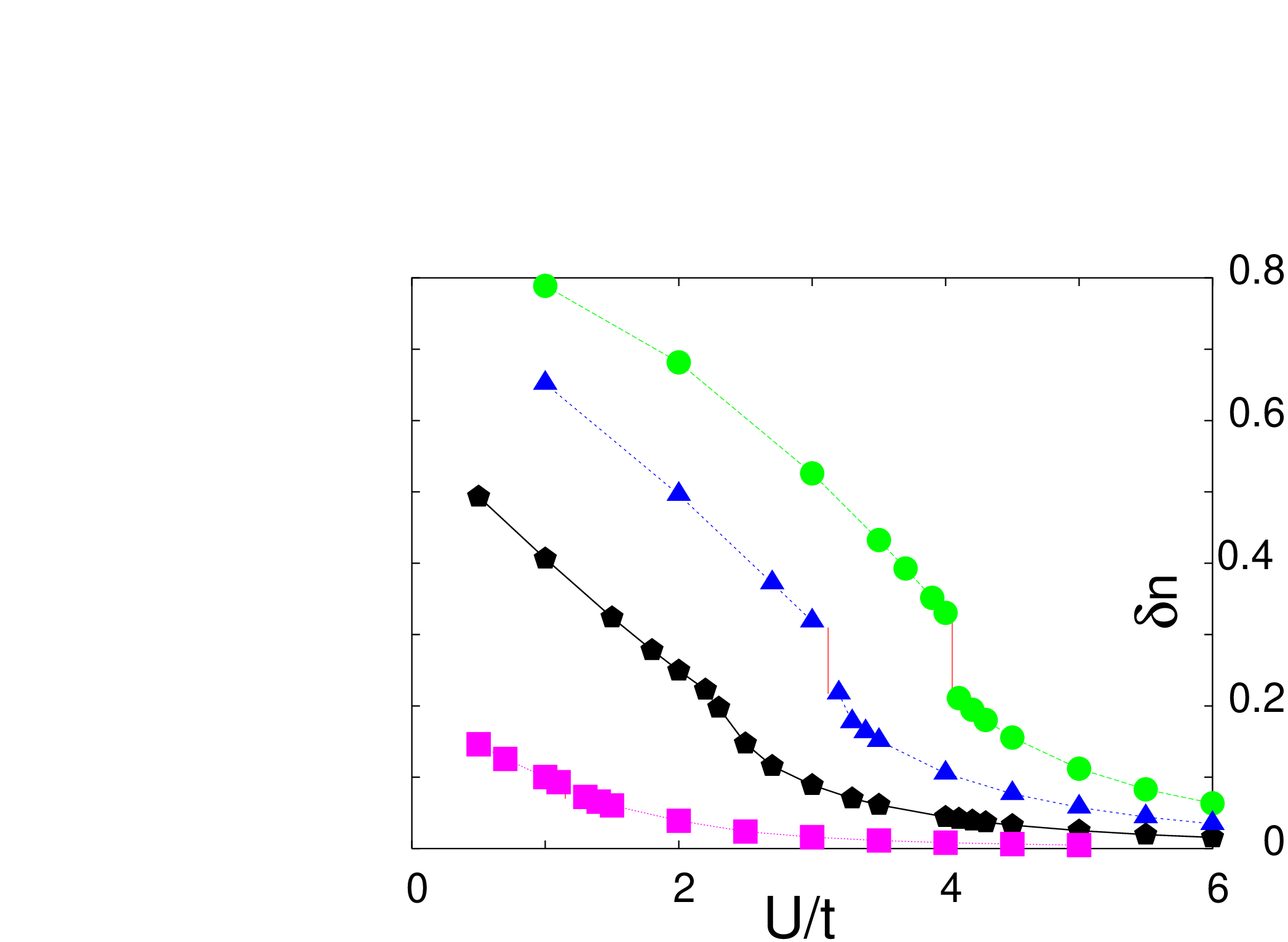}
\caption{Left panel: Staggered magnetization $m_s$ plotted as a function of $U/t$ at half-filling. A first order phase transition takes place with the onset of $m_s$ at $U_{AF}$. Right Panel: Staggered occupancy, i.e., the difference in the filling factor of the two sublattices $\delta n$ plotted as a function of $U/t$ at half-filling. $\delta n$ is non zero for all values of $U/t$. For $U < U_{AF}$, $\delta n$ decreases monotonically and a discontinuity occurs in $\delta n$ at $U_{AF}$.}
\label{m_sc}
\end{center}
\end{figure}

{\bf{Half-Filling:}}
The left panel of Fig.~\ref{m_sc} shows our results for the staggered magnetization $m_s$, defined as $m_s =(m_{zB}-m_{zA})/2$ where $m_{z\alpha} = n_{\ua \alpha}-n_{\da \alpha}$ is the sublattice magnetization. For a given value of $\Delta$, there exists a threshold value $U_{AF}$ at which the staggered magnetisation turns on with a jump resulting in a first order phase transition.  Due to the presence of the staggered potential, the AFM instability does not occur at arbitrarily small $U$, and a finite value of $U$ is required to turn on the magnetisation. Both $U_{AF}$ and the jump in $m_s$ at $U_{AF}$ are increasing functions of $\Delta$.
Note that since at half filling $n_{A\sigma}+n_{B\sigma}=1$, the uniform magnetisation $m_F= n_\ua -n_\da = (m_{zA}+m_{zB})/2 =0$ in the half filled case.
The right panel of Fig.~\ref{m_sc} shows the staggered occupancy, i.e., the difference in the filling factor of the two sublattices, defined as
$\delta n = (n_{B}-n_{A})/2$.  Due to the staggered on site potential, $\delta n$ is always non zero, even though the Hubbard $U$ tries to suppress it. $\delta n$ decreases monotonically as a function of $U$. At $U_{AF}$, $\delta n$ also shows a discontinuity.

Next we discuss the single particle DOS $\rho_{\alpha,\sigma}(\omega) = -\sum_k Im~\hat{G}_{\alpha\sigma}(k,\omega^{+})/\pi$ where $\alpha$ represents the sublattice $A,B$ and $\sigma$ is the spin.
The spectral gap $E_{g\sigma}$ in the single particle DOS $\rho_{\sigma}(\omega)$ is shown in Fig.~\ref{Gap}.  For  $U < U_{AF}$, the spectral gap is same for both the spin components due to the spin symmetry of the paramagnetic BI phase and $E_{g\sigma}$ reduces with increase in $U/t$. This is because the Hubbard $U$ suppresses the effect of the staggered potential $\Delta$, which is responsible for a non-zero gap in the BI phase. At $U=U_{AF}$, there occurs a jump separating the spectral gaps such that $E_{g\da} <  E_{g\ua}$. For $U > U_{AF}$, $E_{g\da}$ keeps decreasing with increase in $U/t$ and vanishes at $U_{HM} > U_{AF}$, while $E_{g\ua}$ starts increasing with increase in $U/t$ and stays non zero at $U_{HM}$. Thus at half-filling we have an HM AFM point at $U=U_{HM}$, details of which will be published elsewhere~\cite{ihm3_AG}.
For $U> U_{HM}$, the system is an AFMI in which both the gaps increase with increase in $U$. The phase diagram on the basis of above analysis is shown in the right panel of Fig.~\ref{Gap}.
%From the low $\omega$ analysis of the spectral function, one can write $E_{g\sigma} = Z_{\sigma}|\Gamma_{\sigma}|$ with $\Gamma_\sigma = \Delta-U/2(\delta n+\sigma m_s) + S_{\sigma}$ with $S_{\sigma} =P\int_{-\infty}^{\infty}d\omega\Sigma_{A\sigma}^{\prime\prime}(\omega)/\pi\omega$ and $1-Z^{-1}_\sigma$ is the coefficient of linear term in the real part of the self energy. Since $|S_\sigma| \ll \Delta$, $U_{AF} \sim U_{HM}$ at which $E_{g\ua}$ vanishes, can be estimated to be $\ge 2\Delta/\delta n \gg \Delta$.
Note that the spectral gaps are different for the up and down spin components (which is a key feature to get the HM phase) in this model because of the presence of the staggered potential $\Delta$. In the next section we discuss the formation of the Ferrimagnetic HM in the doped case.
\begin{figure}
\begin{center}
\hspace{-1.0cm}
\includegraphics[width=2.8in,angle=0]{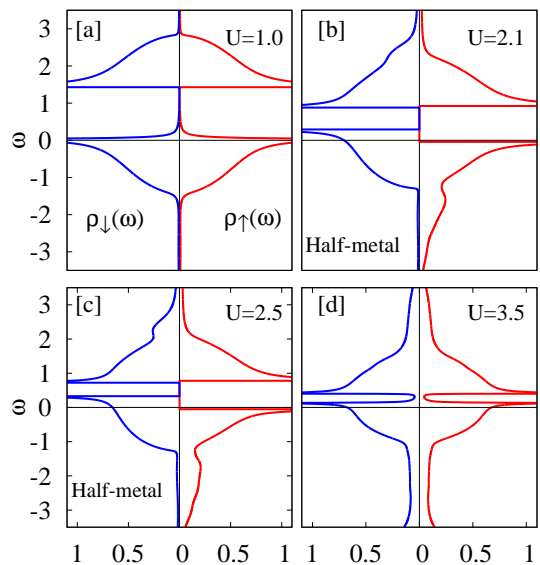}
\caption{Spin-resolved single particle DOS $\rho_{\sigma}(\omega)$ vs $\omega$ for $x=0.17$ and $\Delta=1.0t$. [a] For $U < U_1$, in the metallic phase, the system has spin symmetry with the DOS for both the spin components being non zero at $\omega=0$. For $U>U_1$, e.g. at $U=2.1t$ and $2.5t$, $\rho_{\ua}(\omega=0)=0$ while $\rho_{\da}(\omega=0) \ne 0$. This is the HM phase shown in [b] and [c]. For $U >U_2$, the spin symmetry is restored with $\rho_\sigma(\omega=0)\ne0$ and the system is a regular metal as shown in [d].}
\label{dos_doped}
\end{center}
\vskip-6mm
\end{figure}

{\bf{Doped Case:}} The phase diagram (Fig.~\ref{phase_diag}) in the doped case has a broad Ferrimagnetic HM phase for $U_1 < U < U_2$. 
%In the doped case, for small values of $U$, system is a PM metal (See Fig [1]). As $U$ increases, a first order transition takes place at $U_1$ with the onset of magnetic order and the system becomes a Ferrimagnetic HM. For a wide range of $U$, this phase survives till at $U_2 > U_1$, another first order transition occurs and the system loses its magnetic order. For $U > U_2$, system behaves like a PM metal.
Below we discuss our results in detail explaining how we determine $U_1$ and $U_2$ from an analysis of the single particle DOS and the magnetic properties.

\begin{figure}
\begin{center}
\hspace{-1.0cm}
\includegraphics[width=3.25in,angle=0]{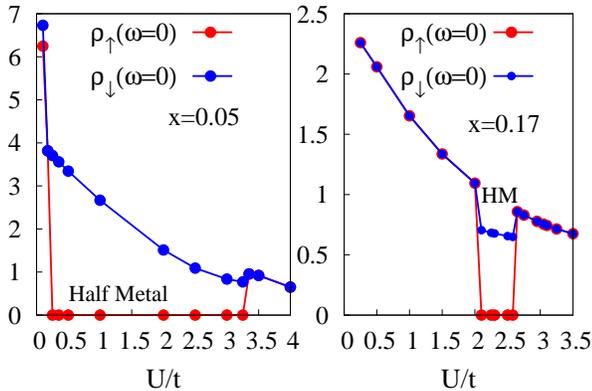}
\caption{Plot of $\rho_\sigma(\omega=0)$ vs $U/t$ for $x=0.05$ and $x=0.17$ at $\Delta=1.0t$. At $x=0.05$ there exists a broad HM phase in which $\rho_\ua(\omega=0)=0$ while $\rho_\da(\omega=0) \ne 0$. The width of the HM phase shrinks in $U$ space as x increases.}
\label{dos_w0}
\end{center}
\vskip-6mm
\end{figure}

{\bf{Single particle DOS in the doped case}}: Fig.~\ref{dos_doped} shows the single particle DOS $\rho_{\sigma}(\omega)=\rho_{A,\sigma}(\omega)+\rho_{B,\sigma}(\omega)$ for both the spin components where $\omega$ is measured from the chemical potential $\mu$.  For $U < U_1$, the DOS for both the spin components is same. In this regime the system is a PM metal since the chemical potential lies inside the lower band for both the spin components (Fig. [4a]). %As $U/t$ increases, the gap in the single particle spectrum decreases.
 For $U > U_1$, magnetic order sets in making the two gaps and the DOS different for the two spin components (Fig. [4b,4c]). The Fermi level lies inside the lower band for the down-spin component making $\rho_{\da}(\omega=0) \ne 0$ while
the up-spin DOS $\rho_{\ua}(\omega=0)=0$ as shown in Fig.~\ref{dos_w0}; hence the system is a HM. For $U>U_2$, the Fermi level lies inside the lower band for both the spin components. This makes both the spin components conducting, with equal density of up and down spins, and the system becomes a paramagnetic metal (Fig. [4d]). Note that there is still a small band gap (at energies higher then $\mu$). At even larger values of $U$, this gap will open up again separating the lower and the upper Hubbard band with the chemical potential being inside the lower Hubbard band.

\begin{figure}[h!]
\begin{center}
%\hspace{-1.0cm}
\includegraphics[width=3.5in,angle=0]{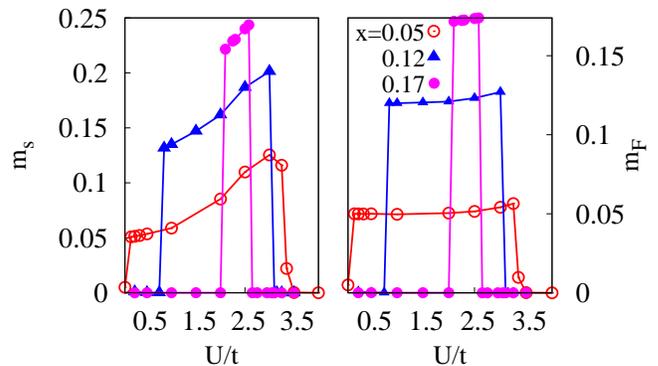}
\caption{The left panel shows the staggered magnetisation $m_s$ vs $U/t$ and the right panel shows the uniform magnetisation $m_F$ vs $U/t$. At $U_1$ there occurs a first order jump in both $m_s$ and $m_F$ making them non zero. Note that in the HM phase, $m_F \sim x $, as explained in the text. At $U_2 > U_1$ there occurs another 1st order transition with $m_s$ and $m_F$ dropping to zero. $\Delta=1.0t$ in both the panels.}
\label{mz_mf_mc}
\end{center}
\end{figure}

{\bf{Magnetisation}}:
The curves for $U_1$ and $U_2$ in Fig.~\ref{phase_diag} are consistent with the magnetic properties as well which are shown in Fig.~\ref{mz_mf_mc}.  For small $U$ values, magnetic order is not favoured. As $U$ increases, a first order transition occurs at $U_1$ when both the sublattices acquire non zero magnetisation $m_{zA}$ and $m_{zB}$ with a jump at $U_1$. Since the system is doped, these magnetisations are not equal and opposite to each other. This results in  a non-zero staggered magnetisation $m_s = (m_{zB}-m_{zA})/2$ as well as a non-zero uniform magnetisation $m_F = (m_{zA}+m_{zB})/2$.
 %As $U$ increases, a first order transition occurs at $U_1$ when both $m_s$ and $m_F$ become non-zero discontinuously.
Within the HM phase $m_F$ and $m_s$ increase with increasing $x$. This is because, in the HM phase, the up-spin band is fully occupied implying $n_{\ua} =1/2$ and all the holes are doped in the down-spin band making $n_{\da}  = n-1/2 $. Therefore, the uniform magnetisation $m_F$, which can also be written as $n_{\ua}-n_{\da}$, goes as $1-n = x$ and is independent of the interaction strength $U/t$. This is in agreement with our results in Fig.~\ref{mz_mf_mc}, within the numerical error-bars.
The staggered magnetisation $m_s$, however, increases with increasing $U/t$.  At $U_2 > U_1$ there occurs another first order transition and the system becomes a paramagnetic metal with both $m_F=m_s=0$ for $U > U_2$.

It is interesting to compare our DMFT phase diagram with the phase diagram within Hartree-Fock (HF) theory, details of which will be published elsewhere~\cite{ihm3_AG}. One can get a HM phase within a HF theory, but it overestimates the tendency to the formation of the HM, and also predicts qualitatively wrong results. Within HF theory for all $U > U_1$, the system is a HM, the reason being the lack of quantum fluctuations in the HF theory which are captured within DMFT.
%Consequently the magnetic order is no longer killed even at very large $U/t$, once it is established.
 \begin{figure}[h!]
\begin{center}
\vskip-1.0cm
%\hspace{-1.0cm}
\includegraphics[width=2.8in,angle=0]{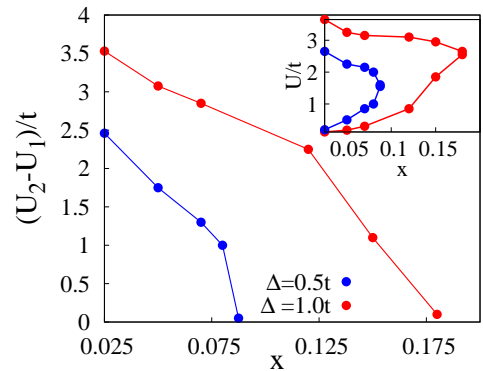}
\vskip-0.1cm
\caption{The width $(U_2-U_1)/t$ of the HM phase in $U$ space as a function of the hole doping $x$. Inset shows the phase boundaries of the HM phase for $\Delta=1.0t$ and $0.5t$. Note that the HM phase gets wider with increases in the staggered potential $\Delta$.}
\label{U2_U1}
\end{center}
\end{figure}

%We end this section with a discussion on the phase boundary of the HM phase for different values of the staggered potential $\Delta$.
%Fig~\ref{U2_U1} shows
%the width $U_2-U_1$ of the HM phase in the $U$ space as a function of hole doping $x$.
We note that, as shown in Fig.~\ref{U2_U1}, the HM phase gets wider with increase in the staggered potential $\Delta$. This will be useful from the application point of view. One should look for correlated band insulators with large bare band gap, and an appropriately larger U, to get a robust HM phase.

{\bf{Conclusions}}:
In conclusion, we have studied a theoretical model in which e-e interactions in a doped band insulator give rise to HM ferrimagnetic phase. Specifically, we studied an extension of the Hubbard model which includes a staggered potential that makes the system a band insulator for $U=0$ at half filling. As we turn on the on-site repulsion $U$, an AFM order sets in with a first order transition at some threshold value of $U = U_{AF}$. The AFM phase has different spectral gaps for the two spin components and on doping away from half filling it becomes a HM ferrimagnet. In the doped system the HM phase survives for $U < U_{AF}$ as well. The width of the HM phase in $x$ space is largest for $U \sim 2\Delta$ and increases with increase in $\Delta$.

We emphasize that the new mechanism for half-metallicity described in
our paper is quite distinct from the mechanisms in well-known materials that exhibit
this phenomenon like the manganites, double-perovskites, or Heusler
alloys, all of which have both local moments and itinerant electrons.
Recently, there have been other theoretical discussions of
half-metallic behavior as well [11]. However, in all of these works
either the HM phase exists only for some special doping values or
require an external electric or magnetic field. The HM phase we
discuss exists for a broad range of doping and is not dependent on
application of external fields.

Although our finding is based on the study of a specific model, we
expect that for any system where the AFM phase at half filling has
different spectral gaps for the two spin components and AFM order
survives on doping, one should expect to see a HM phase. We hope that our study will motivate a search for
materials along this direction and open up new possibilities in the area of spintronics.

M.R. would like to acknowledge support from DOE-BES DE-SC0005035 and
his collaboration with H.R.K. was made possible by NSF MRSEC DMR-0820414. HRK would like to acknowledge support from the DST, India.


\begin{thebibliography}{99}
\vspace{1cm}
\bibitem{HMAFM_expt}
W. E. Pickett and H. Eschrig, J. Phys.: Condens. Matter {\bf{19}}, 315203 (2007); Xiao Hu, Adv. Materials {\bf{24}}, 294 (2012), {\it Half-metallic Alloys: Fundamentals and Applications}, Lecture Notes in Physics, Vol. 676, I. Galanakis, and P.H. Dederichs, Springer (2005).
\bibitem{garg}
A. Garg, H. R. Krishnamurthy, and M. Randeria, Phys. Rev. Lett. {\bf{97}}, 046403 (2006).
\bibitem{ihm-metal}
L.Craco, P. Lombardo, R. Hayn, G. I. Japaridze and E. Muller-Hartmann, Phys. Rev. B {\bf 78}, 075121 (2008); N. Paris, K. Bouadim, F. Herbert, G. G. Batrouni, and R. T. Scalettar, Phys. Rev. Lett. {\bf 98}, 046403 (2007); A. T. Hoang, J. Phys.: Conden. Matt {\bf 22}, 095602 (2010).
\bibitem{note}
The true ground state for the IHM for large U indeed has AFM order and the metallic phase we found is typically overwhelmed by AFMI. However, if we can frustrate AFM order, then the metallic ground state will win.
\bibitem{afm_ihm}
S. S. Kancharla and E. Dagotto, Phys. Rev. Lett. {\bf{98}}, 016402 (2007); K. Byczuk, M. Sekania, W. Hofstetter, and A. P. Kampf, Phys. Rev. B {\bf{79}}, 121103 (2009).
\bibitem{doped}
K. Bouadim, N. Paris, F. Herbert, G. G. Batrouni and R. T. Scalettar, Phys. Rev. B, {\bf{76}}, 085112 (2007)
\bibitem{georges}
A. Georges, G. Kotliar, W. Krauth, and M.J. Rozenberg,  Rev. of Modern Phys. {\bf 68}, 13 (1996)
\bibitem{jarrell}
T. Pruschke, M. Jarrell, and J. K. Freericks, Adv. Phys. {\bf 44}, 187 (1995).
\bibitem{kk}
H. Kajueter and G. Kotliar,
Phys. Rev. Lett. {\bf 77}, 131 (1996).
\bibitem{ihm3_AG}
A. Garg, H. R. Krishnamurthy, and M. Randeria (unpublished).
\bibitem{HM_recent}
R. Nandkishore, G. Chern, and A. V. Chubukov, Phys. Rev. Lett. {\bf 108}, 227204 (2012); Z. Hao, and  O. A. Starykh, Phys. Rev. B {\bf 87}, 161109(R) (2012); J. Yuan, D. Xu, H. Wang, Y. Zhou, J. Gao, and F. Zhang,  arxiv:1302.7123.
\end{thebibliography}
\end{document}